\documentclass{aa}  
\usepackage{amsmath}
\usepackage{graphicx}
\usepackage{graphics}
\usepackage{subfigure}
\usepackage{txfonts}
\usepackage{natbib}
\bibpunct{(}{)}{;}{a}{}{,}

\def\ltsima{$\buildrel<\over\sim$}
\def\lsim{\lower.5ex\hbox{\ltsima}~}
\def\gtsima{$\buildrel>\over\sim$}
\def\gsim{\lower.5ex\hbox{\gtsima}~}

\def\msolyr{M$_{\odot}$~yr$^{-1}$}

\def\teff{\ifmmode T_{\rm eff} \else $T_{\mathrm{eff}}$\fi}

\def\lya{Ly$\alpha$} 
\def\ha{H$\alpha$} 
\def\hb{H$\beta$}

\def\ergscm{erg~s$^{-1}$~cm$^{-2}$}
\def\cm2{cm$^{-2}$}
\def\kms{km~s$^{-1}$}
\def\myr{M$_{\odot}$~yr$_{-1}$}

\def\ewlya{$EW_{\mathrm{Ly}\alpha}$}

\def\hi{H{\sc i}}
\def\hii{H{\sc ii}}

\def\nh{\ifmmode N_{\mathrm{HI}}\else $N_{\mathrm{HI}}$\fi}
\def\nhi{\ifmmode N_{\mathrm{HI}}\else $N_{\mathrm{HI}}$\fi}

\def\izw{{\sc I}Zw 18}
\def\vexp{\ifmmode v_{\rm exp} \else v$_{\rm exp}$\fi}
\def\taua{\ifmmode \tau_{a}\else $\tau_{a}$\fi}

\begin{document}
  \title{On the origin of \lya\ absorption in nearby starbursts and implications for other galaxies\thanks{Based on observations made with the Hubble Space Telescope
  obtained from the ESO/ST-ECF Science Archive Facility} }
  \subtitle{}
  \author{Hakim Atek\inst{1}, Daniel Schaerer\inst{2,3}, Daniel Kunth\inst{1}}
  \institute{Institut d'Astrophysique de Paris (IAP), 98bis boulevard Arago, 75014 Paris, France
         \and Observatoire de Gen\`eve, Universit\'e de Gen\`eve, 51 Ch. des Maillettes, 1290 Sauverny, Switzerland
         \and
Laboratoire d'Astrophysique de Toulouse-Tarbes, 
Universit\'e de Toulouse, CNRS,
14 Avenue E. Belin,
31400 Toulouse, France
 }

\authorrunning{}
\titlerunning{On the origin of \lya\ absorption in starbursts}

\date{Received date; accepted date}

  \abstract
  {Despite the privileged position that Lyman-$\alpha$\ (\lya) emission line holds in the exploration of the distant universe and modern observational cosmology, the origin of the observed diversity of \lya\ profiles remains to be thoroughly explained. Observations of nearby star forming galaxies bring their batch of apparent contradictions between \lya\ emission and their physical parameters, and call for a detailed understanding of the physical
processes at work. \izw, one of the most metal-poor galaxies known is of particular interest
in this context.}
  {Fitting the \lya\ spectrum of \izw\ to understand the origin of the damped absorption profile and its spatial variations across the NW region. To establish a general picture of the physical parameters governing the \lya\ strength and profile both in local and in high-z galaxies.}    
  {We use a 3D \lya\ radiation transfer code to model Hubble Space Telescope (HST) observations of \izw. Observational constraints of relevant parameters such as dust or \hi\ column density are derived from previous studies and from the present analysis. Different geometrical configurations of the source and the neutral gas are explored.}    
  {The integrated \lya\ profile of NW region of \izw\ is reproduced using the observed small amount of dust ($E(B-V) \approx 0.05$) and a spherical \hi\ shell with $\nhi = 6.5\times 10^{21}$ cm$^{-2}$. Such a high column density makes it possible to transform a strong \lya\ emission (\ewlya\ = 60 \AA) into a damped absorption even with a small extinction. When a slab geometry is applied and a given line of sight is chosen, the \lya\ profile can be successfully reproduced with no dust at all and  $\nhi = 3 \times 10^{21}$ cm$^{-2}$. The spatial variations of the profile shape are naturally explained by radiation transfer effects, i.e.\ by scattering of \lya\ photons, when the observed
surface brightness profile of the source is taken into account.
In the case of outflowing Inter Stellar Medium (ISM), as commonly observed in Lyman Break Galaxies (LBGs), a high \nh\ and dust content are required to observe \lya\ in absorption. For nearly static neutral gas as observed in \izw\ and other low luminosity galaxies
only a small amount of dust is required provided a sufficiently high \nh\ covers the galaxy. We also show 
how geometrical and apertures effects affect the \lya\ profile.} 
   {}

  \maketitle

\section{Introduction}
The detection of high redshift ($z$) galaxies has become, through the
last decade, a routine fact, although the discovery of primeval
galaxies that are forming their first stars still remains a
challenge. Depending on the selection techniques, mainly two classes
of galaxies are found: Lyman Break Galaxies (LBGs) selected by their
UV continuum break, and Lyman-Alpha Emitters (LAEs) selected upon
their \lya\ emission line. The situation was, however far different
before this successful era. \citet{pp67} were the first to estimate that
young distant galaxies should be detectable through a strong
\lya\ emission. Nevertheless it took nearly thirty years until such
populations could be found at $z \sim$ 2--7, thanks in particular to
instruments with large field of view (FOV) and 4-8m class telescopes
\citep[see e.g.][]{hu98,kudritzki00,Malhotra02,ajiki03,taniguchi05,Shimasaku06,
 kashikawa06,Tapk06,gronwall07,ouchi08,nilsson08}.  Only recently, GALEX
has provided for the first time a comparable survey at low redshift
\citep[$z \sim$ 0.2--0.35][]{deharveng08}, thanks to its wide FOV and spectroscopic
capabilities in the UV.

Earlier studies of nearby galaxies using mostly the UV capabilities of
IUE and HST recognized quickly that \lya\ emission was fainter than
naively expected from recombination theory and that the \lya\ line
showed a great diversity of profiles from absorption to emission
\citep[e.g.][]{meier81, hartmann84, deharveng86, hartmann88,
 terlevich93,lequeux95,thuan97a,thuan97b,kunth98}.  Later, HST has
allowed one to map the spatial distribution of \lya\ emission and
absorption, of the stellar sources, nebular emission, and dust
\citep[see][]{kunth03,mashesse03,hayes05,hayes08,atek08}.
Despite these information, no clear picture has yet emerged explaining
consistently the \lya\ and related observations in nearby starbursts.

Indeed, it is now well known, both theoretically and empirically that
different physical processes affect the \lya\ intensity, profile
shape, and ``visibility'' (i.e.\ detection frequency among
starbursts)\footnote{See \citet{schaerer07} for an overview.}:
destruction of \lya\ photons by dust \citep[cf.][]{neufeld90,charlot_fall93}, 
velocity fields in the ISM \citep{kunth98,lequeux95}, 
an inhomogeneous ISM \citep{neufeld91,giavalisco96,hansen06}, 
underlying stellar absorption \citep{valls93}, 
and star formation duty cycles \citep{valls93,Malhotra02}. 

Also an ``unifying'' scenario to explain the observed diversity of \lya\ profiles in terms of an evolutionary sequence
of starburst driven super-shells/superwind has been presented by \citet{tenorio99} and has been confronted with local starburst observations by \citet{mashesse03}.
For distant galaxies, \citet{schaerer08,verhamme08} have recently shown -- using radiation transfer models
and empirical constraints -- that \lya\ line profiles of high-$z$
LBGs and LAEs can well be reproduced and that the diversity of \lya\ from emission to 
absorption is mainly due to an increase of the dust content and the \hi\ column density.
Despite this progress, a global picture identifying the main processes
and explaining this diversity also in a quantitative manner is
still missing for \lya\ in local/nearby galaxies. Furthermore
differences between the high and low redshift samples -- if any --
must be understood. We here provide a first step towards these goals
by examining and modeling one of the most metal-poor star forming
galaxies in the local Universe, \izw, and by putting it into context.

\begin{figure}[!htb]
\includegraphics[width=8.5cm,height=10cm]{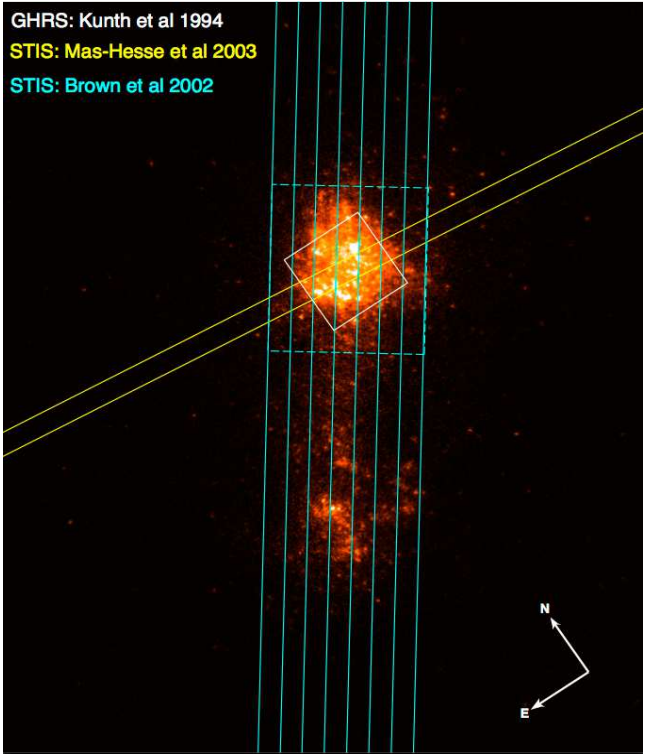}
\caption{FUV Image of \izw\ showing the different observation
  apertures.  The FOV is 16\arcsec $\times$ 18\arcsec. The first slit
  position is in the NE direction and the seventh toward the SW. The
  integration is performed within a region of 4 \arcsec\ along the
  spatial axis of the North-West region and is marked with dashed
  lines.}
  \label{apertures_a}
\end{figure}

Since its discovery by \citet{zwicky66}, \izw\ has been studied 
extensively, and it remains one of the most metal-poor galaxies
known today \citep{skillman93, izotov99}.
Its main \hii\ region (called the NW region, cf.\ Fig.\ \ref{apertures_a})
showing strong optical emission lines is clearly a site of very recent
($<10$ Myr) and/or ongoing massive star formation
\citep[cf.][]{hunter95,demello98,brown02}.
Therefore the finding of a broad damped \lya\ absorption line 
by \citet{kunth94,kunth98} came as a surprise, where strong emission
was predicted, given the strong optical H recombination lines
and the low dust content expected for such low metallicities
\citep[cf.][]{kunth94,terlevich93}.
Observations of SBS 0335-052, nearly as metal-poor as \izw, showed
later a similarly broad profile \citep{thuan97b}.  
However, since compared to \izw,
SBS 0335-052 has a higher extinction
and is now known to
harbour more dust both in absolute terms (dust mass) and in relative
terms ($L_{\rm IR}/L_{\rm UV}$) \citep{thuan99,houck04,wu07,engelbracht08},
it is a priori more challenging to explain \lya\ absorption in \izw\ than
in SBS 0335-052. For these reasons \izw\ represents an ideal
test case to understand how strong intrinsic \lya\ emission is transformed
to the observed broad \lya\ absorption, in a dust poor (but not dust-free), 
very metal-poor galaxy.

\citet{kunth94,kunth98} suggested that both dust absorption and
multiple scattering of \lya\ photons out of their narrow (2.0\arcsec$\times$2.0\arcsec) 
GHRS/HST aperture as the most natural explanation for the observed weakness of \lya\
in \object{\izw}. They also noted that all galaxies showing \lya\ absorption (4/8 in their
small sample) showed nearly static neutral gas, which must increase
the mean free path of \lya\ photons.
However, the IUE spectrum of \izw\ taken with an entrance hole of 
20\arcsec $\times$10\arcsec\ shows basically the same profile, indicating
that at least over 5--10 times larger scales no significant amount of \lya\ 
emission is recovered. In any case, whether quantitatively these explanations
are viable remains to be seen. This is one of the concrete goals of the 
present paper.

To address the above questions we will use the most recent
observations of \izw\ and our state-of-the-art 3D \lya\ and UV continuum
radiation transfer code MCLya \citep{verhamme06}. This will in particular
allow us to carefully examine in a quantitative manner the possible explanations
leading to \lya\ absorption in \izw.  Finally we will also discuss other nearby
starbursts with \lya\ absorption, and place the local objects in a
broader context.

Our paper is structured as follows.
In Sect.\ \ref{observations_sec} we describe the main observations from HST and other
facilities and summarize the main observational constraints. In Sect.\ \ref{analysis_sec}
we set out to explain the \lya\ absorption in \izw, discussing our radiation transfer modeling 
tool, geometrical effects, and presenting modeling results for different ISM geometries.
Our results for \izw\ are discussed and compared to other nearby and high-$z$ starbursts
in Sect.\ \ref{comparison_sec}. In Sect.\ \ref{conclusions_sec} we summarize our main conclusions.

\begin{table}[]
\centering
\begin{minipage}{0.5\textwidth}
\caption{HST observations of \izw. References: (1) \citet{mashesse03}; \ (2) \citet{brown02}; \ (3) \citet{cannon02}  }
\label{data}
\renewcommand{\footnoterule}{}  
\begin{tabular}{l l c c c c }
\hline
\hline
\\
Instrument    &      Filter/    & Band &    Exposure  &    Proposal   &   Ref.  \\
    &  Grating    &  &    time [s] & ID \\
\hline \\
STIS              &   G140M       &\lya\ &  1764          &    GO-8302   &   1             \\
STIS              &   G140L        & \lya&  40360\footnote{Total integration time in the 7 slit positions}      &    GO-9054   &   2                   \\
STIS              &   F25SRF2    &FUV&  5331          &    GO-9054   &   2                   \\
STIS              &   F25QTZ       &NUV&  5786          &    GO-9054   &   2                   \\
WFPC2         &   F487N        & \hb&  2500          &    GO-6536   &   3                   \\
WFPC2         &   F658N        & \ha&  4600        &    GO-5434   &   3                   \\
WFPC2         &   F450W       & B&  4600           &    GO-5434   &   3                   \\
WFPC2         &   F675W       & R&  2000           &    GO-5434   &   3                   \\ 
\\
\hline  
\end{tabular}
\end{minipage}
\end{table}

\begin{figure}[!htb]
\includegraphics[width=9cm,height=7cm]{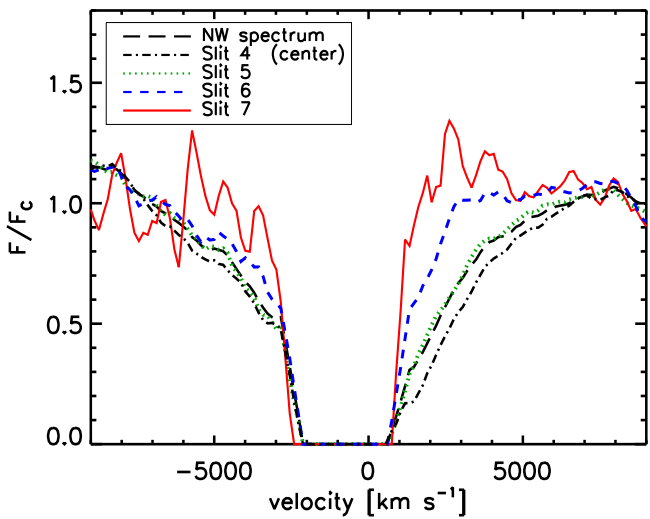}
\caption{STIS spectra of the NW region of \izw\ taken at different
  locations (cf.\ Fig.\ \ref{apertures_a}).  The slit positions 4 to 7
  show the variation of the profile shape from the center to the edge
  of NW region. For comparison, the integrated spectrum in the seven
  slits over the center of NW region is over-plotted. All spectra were
  normalized to the continuum value determined in
  Fig.\ \ref{stis_ghrs}. For the sake of clarity, a cut-off is applied
  on geocoronal emission residuals and spectra are smoothed with a
  3-pixels boxcar.}
   \label{apertures_b}
\end{figure}


\section{Observations} 
\label{observations_sec}
The main observational data used in this paper are summarized in Table \ref{data}.

\subsection{Spectroscopy}
\label{spectroscopy_sec}

We use in this work archival spectroscopic observations obtained, with the Space Telescope Imaging Spectrograph (STIS) onboard HST, under program GO-9054, by \citet{brown02}. G140L grating was used combined with the 52\arcsec $\times$ 0.5\arcsec\ slit. \izw\ was spatially covered with seven adjacent slit positions along its main axis (see Fig. \ref{apertures_a}). Standard calibrations were performed using the CALSTIS pipeline (Ver 2.26), and exposures (two) for each position are registered and co-added. In addition, data were corrected for geocoronal \lya\ emission by fitting and subtracting the nearby background regions in individual spectra. This calibration and spectra extraction were performed using  IRAF and IDL routines.

In Fig.\ \ref{apertures_b} we show the spatial variations of the
\lya\ profile across the NW region. Spectra were extracted from the seven adjacent
positions of the STIS long slit covering the galaxy in the NE-SW axis, providing spatial information in two
directions. Flux was then integrated in each slit along a 4
\arcsec\ aperture centered on the NW component of \izw. 
Finally, an integrated spectrum of the NW region was also constructed
from these integrated slit spectra.
The strength of the \lya\ absorption in these spectra is quantified
by its equivalent width and corresponding \hi\ column density, \nhi,
determined assuming a Voigt profile and $b=20$ \kms\ (cf.\ below). 
These values are reported in Table \ref{lya_variations}.

\begin{figure}
\begin{center}
\includegraphics[width=8.8cm]{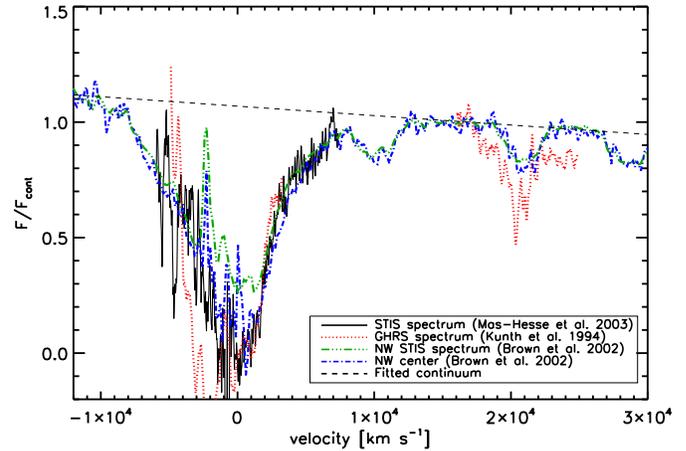}
\caption{\izw\ spectroscopic data.
The figure presents a compilation of spectroscopic informations available for \izw. Geocoronal \lya\ emission has been subtracted from all the spectra. Dark solid line represents the best STIS spectrum around the \lya\ absorption. The red dotted line is the GHRS spectrum which covers a part of the absorption and a part of the UV continuum. The blue dot-dashed line shows the STIS spectrum with a large wavelength coverage ([1100-1750 \AA]) extracted from the center of the NW region. The green long dot-dashed line is the result of an integration over all the slits in the NW region, previously shown in Fig.\ \ref{apertures_b}. It has been used to fit the UV continuum (black dashed line). All the spectra were then normalized to match the continuum value around 1280 \AA\
(i.e.\ at v $\sim$ 16000 \kms). References and legend are given in the inset.}
\label{stis_ghrs}
\end{center}
\end{figure}

Earlier, \lya\ observations of \izw\ were obtained by \citet{kunth94} and
later on by \citet{kunth98} using the Large Science Aperture (LSA,
2\arcsec $\times$ 2\arcsec) of GHRS onboard HST (see
Fig.\ \ref{apertures_a}). \cite{mashesse03} (hereafter MH03) re-observed the galaxy
with better settings using STIS with G140M grating through a 52\arcsec
$\times$ 0.5\arcsec longslit, translating to a spectral resolution
around 0.15\AA\ (37 \kms\ at \lya\ wavelength). The longer wavelength
range of the STIS observations allows a better coverage of the
\lya\ absorption red wing, as compared to GHRS spectrum, and confirms
the large damped \lya\ absorption. In Fig.\ \ref{stis_ghrs}, we plot
together the STIS and GHRS spectra. Due to differences in the
instrument apertures, the spectra had to be matched. 
To fit and estimate the UV continuum (dashed line) we used the archival STIS observations
that allows a broad wavelength coverage to include the absorption wings.
All the spectra were then normalized to the
value where the \lya\ red wing reaches this continuum ($\sim$ 1300\AA). The different \lya\ profiles obtained
are in good agreement. A correct estimation of the continuum around
\lya\ is particularly important for the modeling of the
\lya\ spectral profile (see Sect. \ref{modelisation_sec})

\begin{table}[!htb]
\begin{center}
\begin{minipage}[t]{\columnwidth}
  \renewcommand{\footnoterule}{} 

\caption{Spatial variations of \lya\ properties in STIS slit positions. Columns (2) and (4) indicate respectively the \hi\ column density and the \lya\ line equivalent width derived by fitting each \lya\ absorption with a Voigt profile with $b=20$ \kms. The errors in cols. (3) and (5) are determined from the lower and upper limits of the fits. Last column is the integrated flux in FUV (1500 \AA) image over the NW region and in simulated slits in order to match the aperture used for the extraction of STIS spectra. Same quantities are also given for integrated spectrum in the entire NW region and for the MH03 STIS spectrum.}
\label{lya_variations}
\renewcommand{\footnoterule}{}  
\begin{tabular}{l l c l c c  }
\hline
\hline
\\

Slit   &      \nhi                             &    $\sigma_{N_{H}}$         &   EW$_{Obs}$        &  $\sigma_{EW}$                     &  f(1500 \AA)         \\
         &[cm$^{-2}$]                        &    ($\times 10^{21}$)       &   ~ [\AA]                  &                                                   &   [\ergscm]              \\
\hline \\ 
1     &     1.8 $ \times$10$^{21}$& 0.7                                   &  $-31$                      &  5                                              & 2.8$\times$10$^{-16}$   \\ 
2     &     2.4 $\times$10$^{21}$  & 0.5                                  &  $-32$                     & 3                                             & 9.6$\times$10$^{-16}$       \\
3     &     2.8$\times$10$^{21}$  &   0.5                                 &  $-35$                    &  3                                                & 3.0$\times$10$^{-15}$\\
4     &     2.8$\times$10$^{21}$  &   0.8                                 &  $-34$                    & 5                                                  & 4.1$\times$10$^{-15}$           \\
5     &     2.0$\times$10$^{21}$   &  0.7                                 &  $-30$                    & 5                                               & 2.9$\times$10$^{-15}$    \\ 
6     &     1.0$\times$10$^{21}$    & 0.6                                &  $-20$                    & 6                                               & 1.3$\times$10$^{-15}$      \\
7     &     2.5$\times$10$^{20}$&  0.1                                   & $-10$                     & 3                                                & 5.8$\times$10$^{-16}$          \\
\hline
NW       &  2.1$\times$10$^{21}$ &0.7                                 & $-31$                     & 5                                              & 1.3$\times$10$^{-14}$    \\
\\
MH03   &  2.2$\times$10$^{21}$  &0.7                              & $-30$                        & 4                                              & 2.6$\times$10$^{-15}$  \\
\hline  
\end{tabular}
\end{minipage}
\end{center}
\end{table}

Clearly, \lya\ shows a broad absorption over the entire extent
of the NW region. The width of the profile corresponds to an 
\hi\ column density of \nhi $\sim$ (0.3 -- 3) $\times$
10$^{21}$ \cm2, in agreement with earlier determinations
(\nhi $\sim$ (1.0 -- 3.2) $\times$ 10$^{21}$ \cm2\ from UV observations by 
\citet{kunth98}, although this method does not give systematically the true value of \nh, as we will see later on.
Beyond the scale of the \hii\ NW region ($\sim$ 250 pc), the
UV-optical part of \izw\ is known to be embedded in a large neutral
\hi\ cloud extending over several kpc \citep{vanzee98}.
Furthermore, the strength of the \lya\ absorption decreases clearly
from the center to the border of the NW, as shown in Fig. \ref{apertures_b} for slits 4 to 7 (slits 3 to1 show also a slight decrease),
corresponding to an apparent change of \nhi\ by up to a factor of 
$\sim$ 10. This systematic change of the \lya\ will be explained
below as due to simple radiation transfer effects (Sect.\ \ref{spatial_var_sec}).

Because of the configuration of their apertures, centered on the bright UV peak of the NW region, the GHRS and STIS MH03 spectra are in good agreement (cf. Fig. \ref{stis_ghrs}). This is also true for the central slit of the 2002 STIS data. The integrated spectrum of the NW region shows residual emission at the center that may be due to the contribution of the external slits that show such emission and to geocoronal \lya\ residuals, as the STIS MH03 and GHRS spectra have a better resolution which allow a more reliable correction. Since the profile shape in the wings remains the same, we will take the STIS MH03 spectrum as a proxy for an integrated spectrum of the NW region in our modeling. We derived for this spectrum an \hi\ column density of  $\sim$ 2 $\times$
10$^{21}$ \cm2 from Voigt profile fitting (cf. Table.\ \ref{lya_variations}).

\subsection{Imaging}

UV images, part of the same observing program GO-9054 as that of the STIS spectroscopy, were retrieved from the ESO/ST-ECF 
archive. \izw\ was observed with F25SRF2 filter with bandpass centered at 1457 \AA.  Standard calibrations were performed through CALSTIS pipeline. Images were then corrected for misalignment, divided by the exposure time, and co-added. The final FUV image was then multiplied by PHOTFLAM and PHOTBW header keywords to obtain flux calibrated image.

We also retrieved from the archive HST optical images obtained with the Wide Field Planetary Camera 2 (WFPC2) under programs GO-6536 and GO-5434. Data consist of \ha\ and \hb\ narrow band imaging and corresponding broad band continuum images (see Table \ref{data}). Data were first processed through the standard HST pipeline that gives images in units of counts per second. Multiplying by PHOTFLAM keyword gives fluxes in \ergscm\ \AA $^{-1}$. Finally all images are rotated and aligned to the same orientation and co-added in each filter using inverse variance weighting. OWe estimated the line flux contribution to the continuum images using the appropriate filter throughput ratios at \ha\ and \hb\ wavelengths  and filter width given by PHOTBW. Continuum images were scaled and subtracted from online images, then multiplied by the filter bandwidth to obtain pure emission line fluxes. Continuum subtracted \ha\ and \hb\ images of \izw\ are e.g.\ shown
in \citet{cannon02}. We measured a total \ha\ flux (uncorrected for reddening) of 3.28 $\times$ 10$^{-13}$ \ergscm\  within a circular aperture of 10.5 \arcsec radius, in agreement with values found by \citet{demello98} and \citet{cannon02}.

\begin{figure*}[!htbp]
\centering
\hspace{0.65cm}
\includegraphics[width=6.5cm,height=8cm]{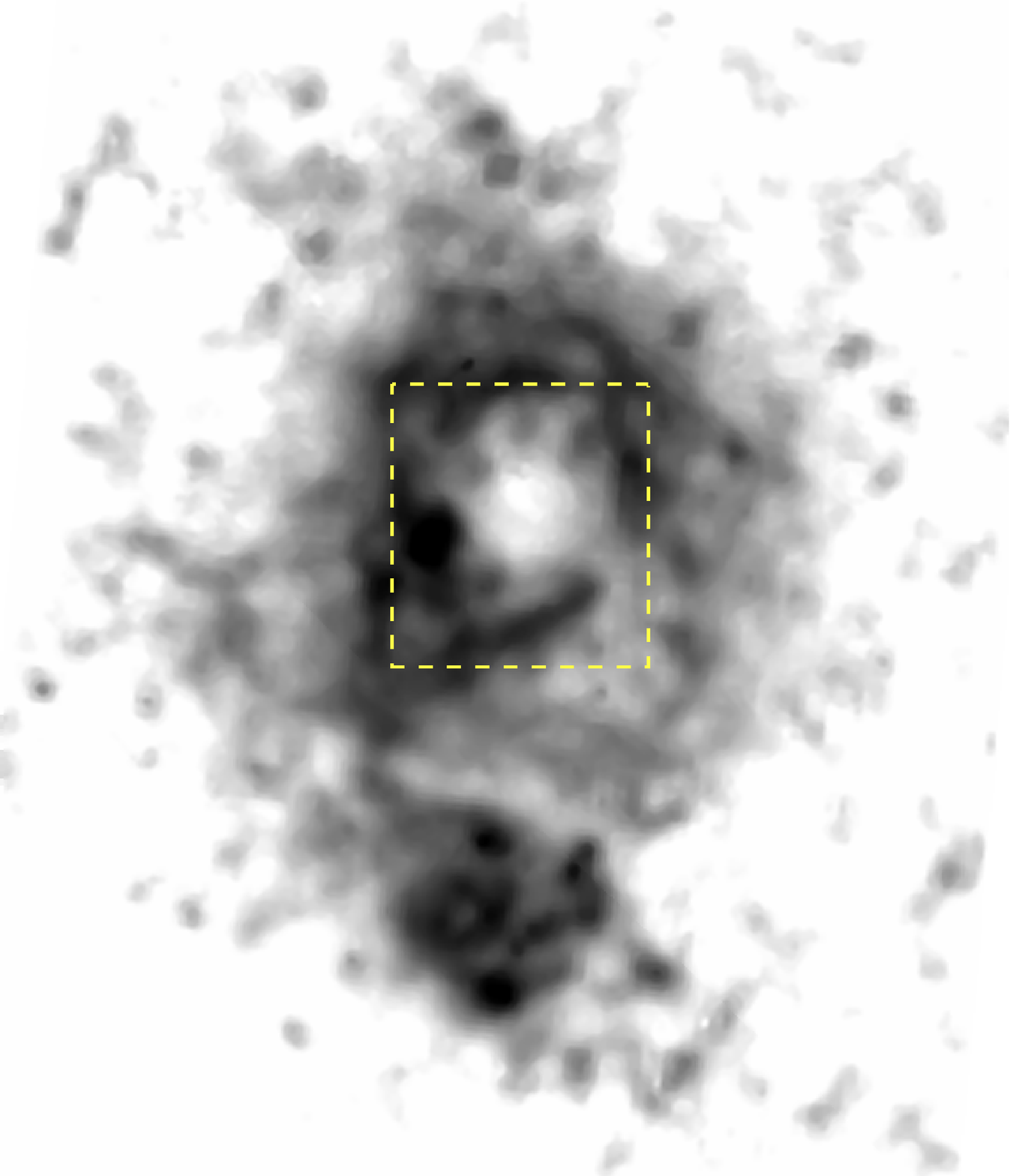}
\hspace{0.4cm}
\includegraphics[width=6.5cm,height=8cm]{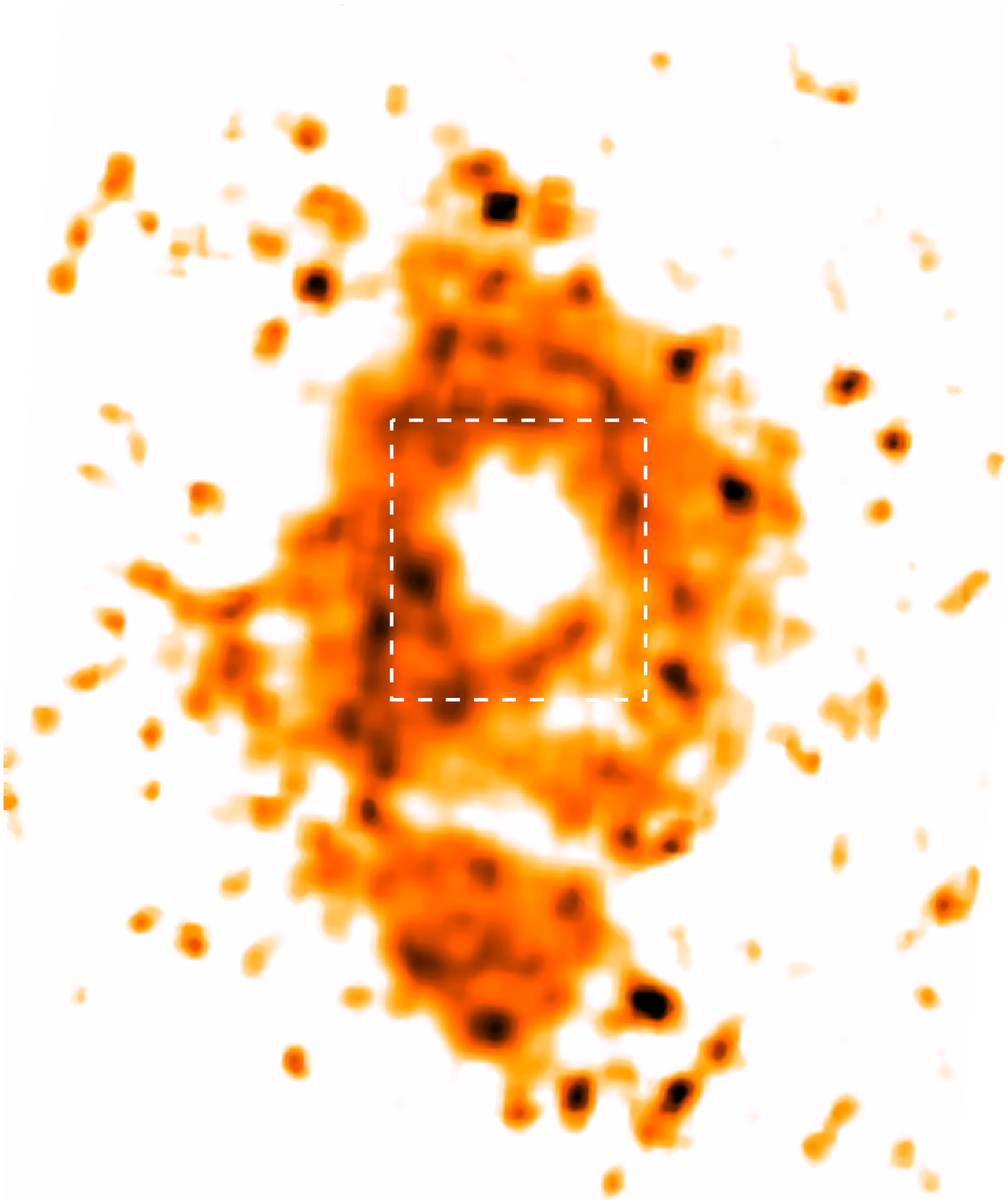}
\vspace{0.3cm} \hspace{0.65cm}
\caption{\izw\ imaging.
\textit{Left:} Intrinsic \lya\ emission map in a logarithmic scale. It has been obtained by correcting the observed \ha\ image with the extinction map and assuming case B recombination theory (see text for details). The result has been smoothed using a median filter (width=5). \textit{Right:} shows the extinction map, in linear scale, obtained from the Balmer decrement \ha/\hb\ then median filtered (width=5). Inverted color scale is used showing higher emission and dust content in darker color. The NW integration box (cf. Fig.\ \ref{apertures_a}) is also shown in dashed line. The size of the field of view is about 13\arcsec $\times$ 15\arcsec\ and the orientation is the same as in Fig.\ \ref{apertures_a}.  } 
\label{lya_maps}
\end{figure*}

{\bf Extinction}
The extinction map of \izw\ is created using the ratio between \ha\ and \hb\ images. In the absence of dust extinction the theoretical value of the Balmer ratio is known to be loosely sensitive to temperature and density. Following \citet{cannon02} we adopted a value of \ha/\hb = 2.76. Potential sources of error on the expected value, such collisional excitation of \hi\ or underlying stellar absorption, are also addressed in this paper.
An E(B-V) map was thus constructed using the following relation:
\begin{equation}
 E(B-V)_{H\alpha/H\beta} = \frac{2.5 \times log(2.76 / R_{obs})}{k(\lambda_{\alpha}) - k(\lambda_{\beta})}
\end{equation}
where $R_{obs} = f_{H\alpha}/f_{H\beta}$\ is the absolute observed flux ratio, and k($\lambda_{\alpha}$), k($\lambda_{\beta}$) are the extinction curve values at H$\alpha$ and H$\beta$ wavelengths respectively. 
 We adopt $k(\lambda_{\alpha}) - k(\lambda_{\beta}) = -1.08$, from \citet{cardelli89}, which is appropriate
for resolved galaxies and diffuse interstellar regions.
Finally, the extinction value was corrected for galactic contribution following \citet{Schlegel98}, which accounts for 0.032 mag. 
The resulting extinction map is shown in Fig.\ \ref{lya_maps}.

Overall the extinction in \izw\ and in its NW region is known to be very low. \citet{mashesse90} found that the Balmer decrement of the whole NW region is consistent with no extinction. However, \citet{dufour88} reported an extinction of $E(B-V) \sim 0.17$ in their 2.5\arcsec $\times$ 6\arcsec\ slit. Ground-based spectroscopic observations revealed typical values ranging from $E(B-V) \sim 0.03$ up to 0.2 \citep[e.g.][]{vilchez98,izotov97a,martin96}. The main reasons of such discrepancies may be differences in the aperture size and the location of the slits on the galaxy, as the dust does not seem to be homogeneously distributed in \izw\ (see Fig.\ \ref{lya_maps}). In the present work, we define a circular aperture (3.2\arcsec\ radius) centered on the NW region. We exclude (1\arcsec\ circular aperture) the central region, where \ha\ and \hb\ emission are much weaker and Balmer ratio gives unreasonably low values, from our measurement. 
The mean color excess derived in this way is $E(B-V) \approx 0.042$.
This value agrees with the determinations by \citet{cannon02} ($E(B-V) =$ 0 -- 0.09) obtained in different parts of the NW region, 
and with \citet{pequignot08}.
We also find no extinction when the central region is not excluded, in agreement with \citet{mashesse90}. 
Subsequently we will adopt an average value of $E(B-V) = 0.05$ for the NW region.

{\bf  Intrinsic \lya\ emission}
From the \ha\ image, and using the extinction map, we have produced a theoretical \lya\ emission map (Fig.\ \ref{lya_maps}).
It is determined using 
$f(Ly\alpha) = 8.7 \times f(H\alpha) \times 10^{(1.048\times E(B-V))} $,  where we assume a case B recombination theory \citep{brocklehurst71} and the extinction law previously mentioned. Naturally, given the small extinction corrections, the 
resulting map of intrinsic \lya\ emission shows a very similar
morphology as the \ha\ map. Schematically, the NW region is 
surrounded by a \lya\ emission shell including in particular
one bright knot \citep[NW1 in the notation of][]{cannon02}). 

\begin{figure}[htbp]
\begin{center}
\includegraphics[width=9cm,height=6.5cm] {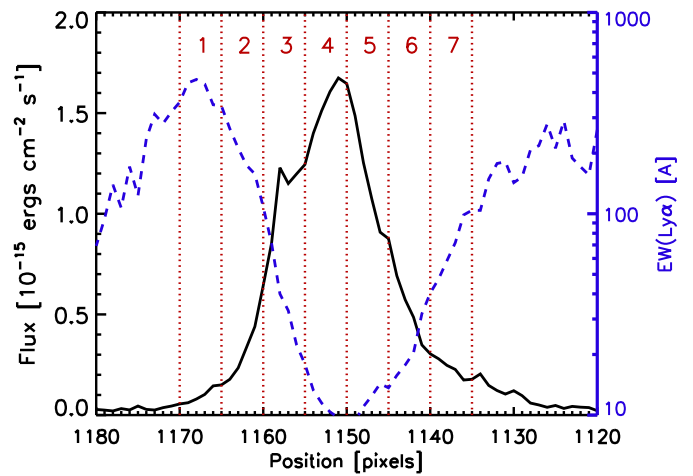}
\caption{ 1D emission profile of NW region. The different profiles are obtained by adding the flux along the slit (spatial direction) in a same aperture used for the extraction of the spectra (cf.\ Fig.\ \ref{apertures_b}) centered on the NW region. Then the 1D profile corresponds to the direction along the dispersion axis. One pixel corresponds to $\sim$ 0.1\arcsec. The dark solid curve is the 1500 \AA\ flux represented on the left y-axis, and blue dashed line the intrinsic \lya\ equivalent width represented in logarithmic scale on the right axis. The position of the seven adjacent  slits are also marked, showing the spatial variation of the emission between the different slits. } 
\label{profile_1d}
\end{center}
\end{figure}

Finally, the spatial variation of the UV continuum, and the intrinsic \lya\ equivalent width are shown in Fig.\ \ref{profile_1d}. The UV image has been corrected for reddening, scaled to \lya\ wavelength using a UV slope of $\beta \sim -2$, and used together with the intrinsic \lya\ map to construct the theoretical \ewlya\ map. Over the entire NW region we obtain \ewlya\ $\sim$ 50 \AA, compatible with expectations for a young starburst. However, as shown in Fig.\ \ref{profile_1d}, we observe very high values around the UV-bright central region. 

Taken together the observations of strong \lya\ {\em absorption}
across the entire NW region despite the presence of intrinsic
{\em strong \lya\ emission} and a very {\em low amount of extinction}
clearly call for a physically consistent explanation
of these apparent contradicting phenomena.

\subsection{Other observational constraints}
\label{obs_constraints}
A mean velocity offset, $\Delta v$(em - abs), between
the systemic velocity, measured from the optical lines, and 
metallic absorption lines of O~{\sc i} and Si~{\sc ii}, was measured by \citet{kunth98}
in the small GHRS aperture centered on the NW region.
They found $\Delta v$(em - abs) $\sim$ 25 \kms, indicating that the neutral gas
is mostly static with respect to the central \hii\ region. 
Recent FUSE observations including other ISM absorption lines
confirm the absence of an outflow in \izw\ on a large aperture
including by far all the UV emitting regions of this galaxy;
\citet{grimes08} measure velocity shifts
between $\sim$ 0 and 40 \kms\ with a mean offset of 8 \kms.

The Doppler parameter $b$ describes the thermal motion of
hydrogen atoms. \hi\ velocity dispersion observed by \citet{vanzee98}
is about 12 - 14 \kms, which translates to $b \simeq 17 - 20$ \kms. A
slightly higher value ($b \simeq$ 27 \kms) was quoted by
\citet{kunth94} from their VLA observations. Given the very damped
profile of the \lya\ absorption, variations within this range of
values does not affect the model fit.

 The Full Width at Half Maximum (FWHM) of the \lya\ emission line can be constrained using FWHM(\ha). \citet{dufour88} found FWHM(\ha) $\sim$ 6.1 \AA\ (280 \kms) from their spectrophotometry observations but with a resolution of 275 \kms. Observations with a better resolution (R $\sim$ 11 \kms\ FWHM) indicates FWHM(\ha) $\sim$ 150 \kms\ \citep{martin96}.  This is consistent with a relatively narrow emission line and we will adopt FWHM(\lya) =100 \kms, although our results are insensitive to the differences found in the observations.

\section{Explaining the \lya\ absorption in \izw}
\label{analysis_sec}

\subsection{General considerations}
\label{analysis_gen_sec}

\begin{figure}[!htbp]
\centering
\includegraphics[width=8.8cm]{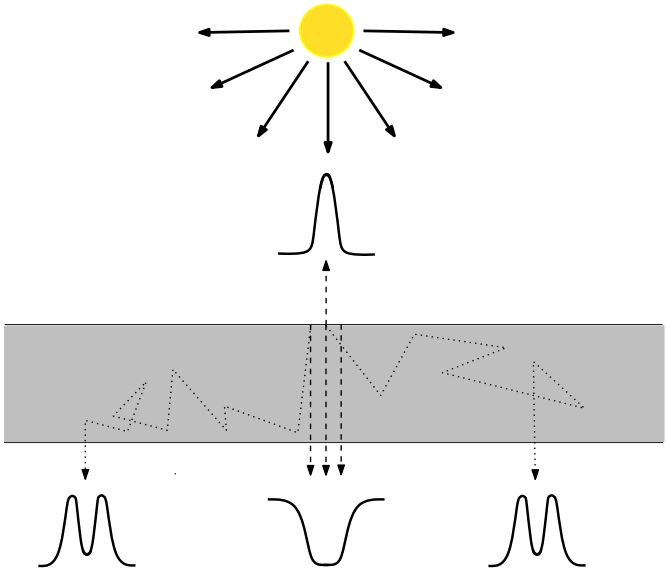}
\caption{Sketch showing geometrical effects on the \lya\ profile shape
  for a point source behind a homogeneous dust-free slab emitting
  pure UV continuum radiation (flat continuum around \lya). The
  output spectra represent the observed profiles in different regions
  and in the observer's line of sight perpendicular to the slab. The
  reflected spectrum by the slab is also represented.  } 
\label{scheme1}
\end{figure}

To transform the intrinsic \lya\ emission (emitted in the \hii\ region) to a
pure absorption profile can in principle only be achieved in two ways:
{\em 1)} by true destruction of \lya\ photons (by dust or possibly by
conversion to two-photon continuous emission in the ionised region), or
{\em 2)} by geometrical effects leading to the scattering of \lya\ photons
out of the line of sight, or by a combination of both.

Examples of line profiles due to dust absorption are shown e.g.\
in \citet{verhamme06,verhamme08} and \cite{schaerer08}. Effect 2) is illustrated
in Fig.\ \ref{scheme1}, showing how for example even a dust-free slab
produces an absorption (Voigt) profile along the central line of sight
from a point-like background source.
This geometrical situation also corresponds to the ``classical'' case
of damped \lya\ systems (DLA) in front of distant quasars, or other \lya\
forest observations.
If the scattering foreground layer was truly dust-free, it is clear that
the photons are conserved; hence the photons scattered away from line center
(causing the apparent absorption line) must emerge somewhere. In a static
configuration, radiation transfer effects redistribute the photons into
the wings, leading to a symmetric double peak \lya\ profile \citep{neufeld90},
as sketched in Fig.\ \ref{scheme1} for the distant, non-central lines
of sight.
Adding dust to effect 2), i.e.\ combining 1) and 2), will reduce the strength 
of the scattered component and further increase the depth of the central 
absorption profile.

Using radiation transfer models we will now examine whether these effects 
can quantitatively explain the observations of \izw, and which of these effects
is dominant.

\subsection{\lya\ and UV continuum radiation transfer modeling}  
\label{modelisation_sec}

\subsubsection{MCLya code and input parameters}

We use an improved version of the Monte Carlo radiation transfer code MCLya of
\citet{verhamme06} including the detailed physics of \lya\ line and UV continuum transfer, 
dust scattering, and dust absorption for arbitrary 3D geometries and velocity fields.
The following improvements have been included \citep[see][for more details]{hayes09}:
angular redistribution functions taking quantum mechanical results for \lya\ into account 
\citep[cf.][]{dijkstra08,stenflo80},
frequency changes of \lya\ photons due to the recoil effect \citep[e.g.][]{zheng02},
the presence of deuterium 
\citep[assuming a canonical abundance of $D/H = 3 \times 10^{-5}$][]{dijkstra06a}, 
and anisotropic dust scattering using the Henyey-Greenstein phase function
(using parameters adopted in \citet{witt00}).
Furthermore a relatively minor bug in the angular redistribution of \lya\ photons
has been fixed, and the code has been parallelized for efficient use on supercomputers.
For the physical conditions in the simulations used for the present paper,
these improvements lead only to minor changes with respect to the MCLya version
used by \citet{schaerer08} and \citet{verhamme08}.
More details on the code upgrade will be given in \citet{hayes09}.

For simplicity, and given the available observational constraints,
all simulations carried out subsequently assume a homogeneous and co-spatial 
distribution of neutral hydrogen and dust with a constant density and temperature.
The corresponding microscopic \hi\ velocity distribution is described 
by the Doppler parameter $b$.
The remaining input parameters of the code are the \hi\ geometry and velocity field,
the spatial location and distribution of the UV continuum and line emission source(s),
and the dust-to-gas ratio.

We consider the following \hi\ geometries:
spherically symmetric shells with a central source,
and plane parallel slabs with a background or internal source (including different
source geometries).
These cases are described by 3 additional parameters:
{\em (i)}   the expansion velocity of the shell, \vexp, or the velocity of the slab with
respect to the source,
{\em (ii)}  the \hi\ column density towards the source, \nhi, and
{\em (iii)} the dust absorption optical depth \taua\ which expresses the dust-to-gas ratio.
As discussed by \citet{verhamme06} \taua\ is related to the usual color excess
$E(B-V)$ by  $E(B-V) \approx (0.06...0.11) \taua$; we assume
$E(B-V) = 0.1$ \taua\ for convenience. 
In short, for a given geometry we have 4 parameters ($b$, \vexp, \nhi, \taua);
$b=20$ \kms\ and $\vexp \approx 0$ \kms\ are constrained by the observations 
(see Sect.\ \ref{obs_constraints}),
\taua\ is varied between 0 (no dust) and 0.5,
the maximum allowed by the observations (Sect.\ \ref{obs_constraints}),
and \nhi\ is varied to reproduce the observed \lya\ line profile.

For each parameter set a full Monte Carlo simulation is run allowing for
sufficient statistics to compute both integrated and spatially resolved
spectra in the \lya\ region. The radiation transfer calculations
cover a sufficiently broad spectral range (here typically from $-10000$ to $+10000$ \kms)
necessary to reach the continuum for the highest column density simulations.
As described in \citet{verhamme06} our MC simulations are computed 
for a flat input spectrum, keeping track of the necessary information
to recompute {\em a posteriori} simulations for arbitrary input spectra.
For the \lya\ fits we assume an input spectrum given by a flat (stellar) continuum
plus the \lya\ line, described by a Gaussian with variable equivalent width \ewlya\ and 
full width at half maximum $FWHM$(\lya).
\ewlya\ is kept free, although constraints are available from our theoretical
(intrinsic) \lya\ map; a $FWHM=100$ \kms\ is assumed as for \ha, although our
results are basically independent of its exact value.
Other continua, such as synthetic high resolution starburst spectra
from \citet{schaerer08}, can also be used.

\subsubsection{Shell models}
\label{shell_sec}

To consider a 
simple geometry to understand the observed \lya\ absorption of \izw, we examine predictions for the {\em integrated
spectrum} of a spherical shell with/without dust. In this case no ``loss'' of 
photons by spatial diffusion is allowed. Hence to transform 
intrinsic \lya\ emission into an absorption profile requires absorption by dust.
We will now examine whether spatially integrated shells can recover the 
observed profile for reasonable amounts of dust and reasonable \hi\
columns. 

\begin{figure}[htbp]
\begin{center}
\includegraphics[width=9cm,height=7cm] {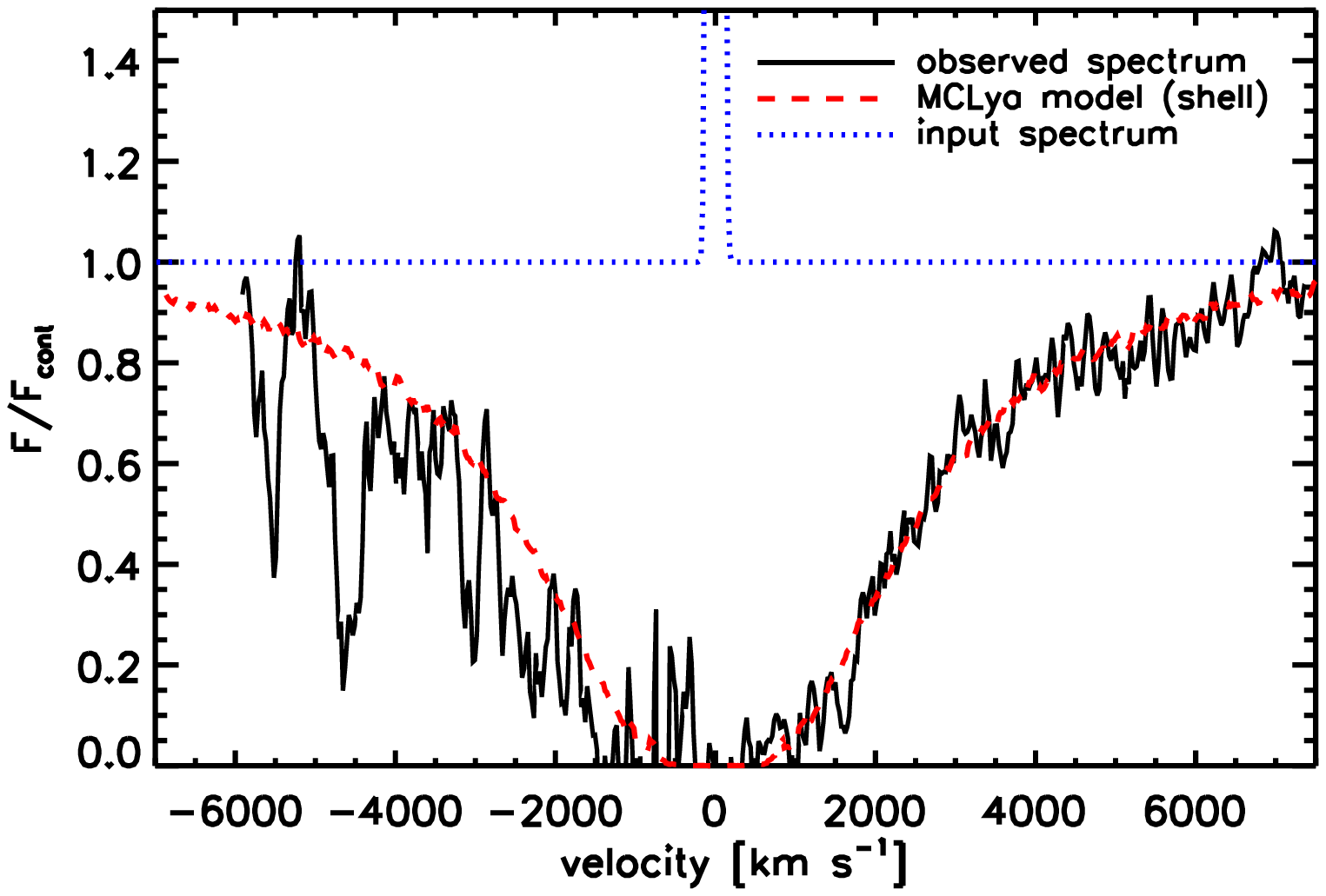}
\caption{Comparison of the observed and fitted \lya\ profile of \izw\ assuming a spherical shell model.
The observed STIS spectrum (from MH03) is represented by the dark line. The model fit, using a shell geometry, is plotted in red dashed line and the parameters used are: \nhi\ = 6.5 $\times$ 10$^{21}$cm$^{-2}$, $\tau$ = 0.5, v=0 \kms, b = 20 \kms. The blue dotted line represents the input spectrum of the simulation. It consists of a flat UV continuum plus a Gaussian \lya\ emission line with $FWHM$=100 \kms. The intrinsic equivalent width adopted is $EW$(\lya)=60 \AA.   } 
\label{lya_fit}
\end{center}
\end{figure}

Adopting an average extinction of $E(B-V) \approx 0.05$ (i.e.\ $\taua=0.5$)
and $b=20$ \kms\ we have computed several static shell models with varying \nhi.
As shown in Fig.\ \ref{lya_fit}, the predicted profile agrees well with the
observations for $\nhi = 6.5\times 10^{21}$ cm$^{-2}$
and for an input spectrum with a \lya\ line equivalent width $EW$(\lya) = 60 \AA. Note that we do not fit the absorptions in the blue wing, attributed to 
Si~{\sc ii} $\lambda\lambda$ 1193.3,1194,5 and  Si~{\sc iii} $\lambda\lambda$ 
1206.5,1207.5 \citep{schaerer08} and possibly to Galactic and intergalactic \hi\ 
absorption, since these are not taken into account in our model.
%

The reason for the resulting  broad damped \lya\ absorption is as follows:
Due to the high \hi\ column density, even a small amount of dust destroys almost all
photons in and around the \lya\ line center. Scattering on hydrogen
atoms with such a high column density, greatly increases the mean path of
\lya\ photons, and hence the probability to be absorbed by
dust. Therefore, the net absorption is only caused by dust absorption,
since, in the present case, we observe all the scattered photons
escaping from the shell, without any line-of-sight effect.

\paragraph{\bf Influence of V$_{exp}$:}
We adopted a static shell in our model to fit the \lya\ absorption
profile. As discussed in \citet{verhamme06}, for increasing \vexp\,
more \lya\ photons will escape from the red part of the line, because
\lya\ photons are seen already redshifted by hydrogen atoms. However,
since the high column density reduces the escape
probability, we can vary the expansion velocity in a certain range
without affecting the quality of the fit. The highest velocity allowed
is around 50 \kms, which already exceeds the observed outflow velocity
of \izw.

\paragraph{\bf Influence of \ewlya:} 
We can show that \ewlya\ close to the maximum value expected by
synthesis models \citep{schaerer03}, for normal IMF populations, are
allowed in our model to fit \izw\ profile. Indeed, when we use an
input \lya\ line with \ewlya\ = 200 \AA\ the fit remains globally
unchanged. Again, in spite the large damped absorption, high
intrinsic \ewlya\ is not excluded by radiation transfer simulations
because of the high \hi\ column density.

\paragraph{\bf Other solutions:} 

We need to invoke a relatively high column density (\nh\ = 6.5 10$^{21}$
cm$^{-2}$) 
to obtain a good fit of the absorption wings. On the
other hand, all photons which, in reality, will scatter away
from the observer's line of sight, are recovered in our simulation,
since we integrate over the entire surface of the shell. 
Relaxing this assumption, i.e.\ considering different geometries, 
would in particular also allow us to lower \nh.

The solution proposed here to fit \izw\ profile is not unique, and different combinations of \nh\ and \taua\  can reproduce the absorption. For instance, the use of a higher value for the extinction (\taua = 1) and a lower \hi\ column density (\nhi = 5$\times$ 10$^{21}$ cm$^{-2}$) produce the same fit quality. 
Overall, this somewhat academic case of a shell model for \izw\
serves to show that even low dust quantities may suffice to transform
\lya\ emission into a broad absorption profile, provided
a sufficiently high column density and/or sufficiently low
outflow velocity, as also discussed in Sect.\ \ref{comparison_sec}.
In any case, the radio observations of \izw\ show very 
clearly a large spatial extension of \hi\ compared to the size
of the NW region (and to that of the spectroscopic apertures).
The effect of such geometries on \lya\ will be addressed now.

\subsubsection{Extended geometries and line of sight effects}
\label{sight_line_sec}


 The galaxy spectrum we observe in reality, could deviate
significantly from the simple homogeneous shell model presented here,
since the source is spatially resolved and the spectrum is not integrated over 
the whole shell surface. Furthermore the spectrum can depend on viewing
angle and on the geometry of the ISM.

\begin{figure}[htbp]
\begin{center}
\includegraphics[width=9cm,height=7cm] {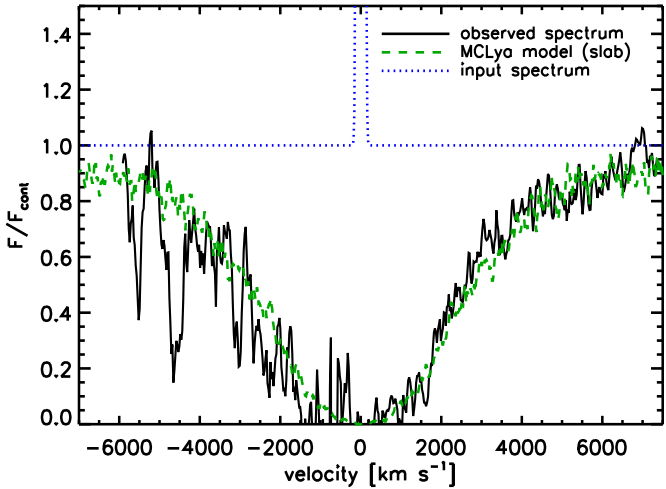}
\caption{ \lya\ absorption fitting {\sc II}. The observed spectrum (STIS MH03) is represented by the dark line. A slab geometry is used for the model spectrum and only photons in the observer's sight-line are collected. It is plotted in green dashed line and the parameters used are: \nhi\ = 3 $\times$ 10$^{21}$cm$^{-2}$, $\tau$ = 0 (no dust), v=0 \kms, b = 20 \kms. The blue dotted line represents the input spectrum of the simulation. It consists of a flat UV continuum plus a Gaussian \lya\ emission line with $FWHM$=100 \kms. The intrinsic equivalent width adopted is $EW$(\lya)=60 \AA.} 
\label{lya_fit2}
\end{center}
\end{figure}

We show in Fig.\ \ref{lya_fit2} that \izw\ absorption can be well
adjusted with lower \hi\ column density than required for the shell
model and without any dust ($E(B-V) = 0$). This result is achieved by
taking a slab geometry with a static gas and applying sight-line
selection criterium, where only photons in the observer's direction
are collected. Then the absorption is caused, not by dust destruction,
but by diffusion of the photons out of the observer's
direction. Strictly speaking, no photon is destroyed. This
demonstrates even better that \lya\ absorption can be observed in
dust-free galaxies (cf. Fig.\ \ref{scheme1} for a schematic
overview). Only a nearly static neutral ISM is required, with \nhi\ $\sim$ 3 $\times$ 10$^{21}$cm$^{-2}$ in this case.

 In this case, we expect to recover the diffused photons
in other directions and/or further from the source.
On the other hand, in presence of dust, this diffuse part would be attenuated or suppressed.
For example, for models with homogeneous gas and dust distributions, our \lya\ transfer
simulations \citep[see][]{hayes09} predict already quite low escape fractions for \lya\ line 
photons, with $f_{\rm esc}$ of the
order of typically 5--10 \% for column densities $\nh \ga 10^{21}$ \cm2, dust optical depths 
$\taua = 0.2$, and low expansion velocities ($\vexp \la 50$ \kms). Much lower
escape fractions ($f_{\rm esc} \sim 10^{-3 \ldots -4}$) are predicted for larger amounts of dust,
such as for the average value adopted for the NW region.
Therefore we expect relatively small amounts of diffuse emission from \lya\ line photons.

\subsubsection{Spatial variations of \lya\ profile}       
\label{spatial_var_sec}

In Sect. \ref{spectroscopy_sec} (Fig.\ \ref{apertures_b}) we have shown that the 
\lya\ profile shows spatial variations between the different STIS slits. 
We now demonstrate, that given the
observational constraints, the \lya\ radiation transport explain
fairly well these variations. We consider a large, static and uniform cloud of
\hi\, (\nhi\ = 3 $\times$ 10$^{21}$ cm$^{-2}$, $b$ = 20 \kms, \taua = 0.5) represented by a slab geometry, covering the NW star-forming
region. We then simulate the observed spatial variations of the
emission strength by using weighted point sources located in front of the \hi\ slab, emitting a flat UV continuum, as input to our
radiation transfer code following the observed UV profile of Fig.\ \ref{profile_1d}. 
The addition of \lya\ line emission will be discussed below.

\begin{figure}[tb]
\begin{center}
\includegraphics[width=8.5cm,height=7cm] {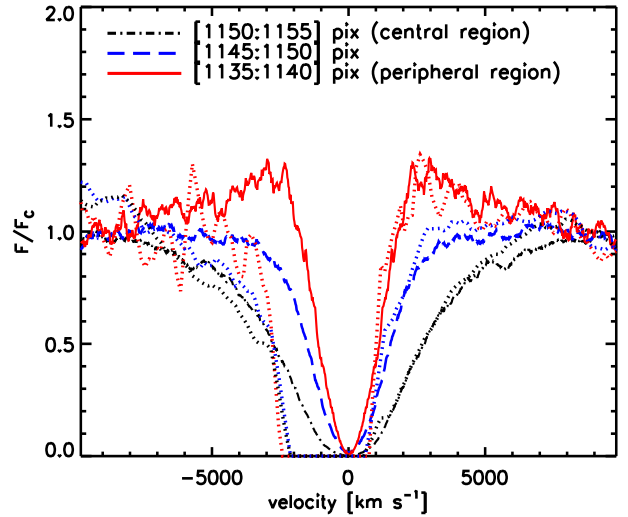}
\caption{Predicted spatial variations of the \lya\ profile. The
  simulation consists of an homogeneous \hi\ slab with \nhi\ = 3
  $\times$ 10$^{21}$ cm$^{-2}$ and \taua\ = 0.5 illuminated with an isotropic, extended, UV source emitting a flat
  continuum. The emission strength is spatially varying from the
  center to the edge to reproduce the observed UV surface brightness
  of \izw. An observer line-of-sight perpendicular to the cube is
  chosen ($\theta$ = 0). The different plotted profiles correspond
  then to different regions at the surface of the slab and are marked
  in units of pixels corresponding to the position of
  the slits in Fig.\ \ref{profile_1d}. The observed profiles (cf. Fig.\ \ref{apertures_b}) are plotted in dotted lines. The blue wing of the profiles is not well fitted because it is affected by geocoronal emission and \hi\ Galactic absorption which the model does not account for.}
\label{spatial_variation}
\end{center}
\end{figure}

The result of this simulation is shown in
Fig.\ \ref{spatial_variation}. The output spectrum is what an observer
would see when his line-of-sight is perpendicular to the slab
surface. At the center (1150 $\le$ pixel $\le$ 1155), in the direction of the brightest
source, we observe the strongest (largest) \lya\ absorption
profile. The profile proves increasingly narrower as one moves away
from the center, what reproduces the trend observed in \izw. The
double-peak contribution, characteristic of diffused photons, can even
be seen in the peripheral region (1135 $\le$ pixel $\le$ 1140).

\begin{figure}[tb]
\centering
\includegraphics[width=8.8cm]{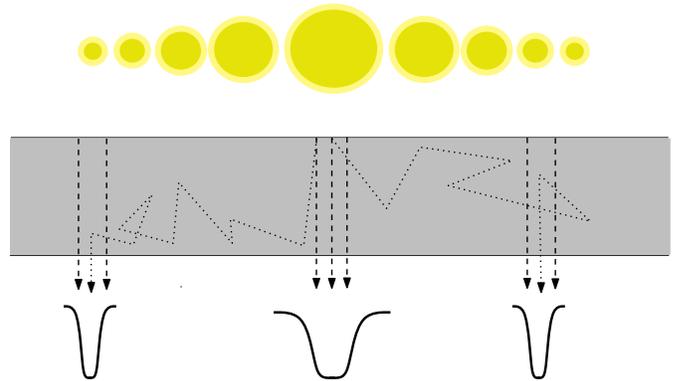}
\caption{Geometrical effects on the \lya\ profile shape: Homogeneous, static, and dust-free slab of neutral gas illuminated by a series of isotropic point sources emitting UV continuum radiation centered on \lya\ wavelength. An extended source is simulated in this case with a varying emission strength, symbolized by the size of the individual point sources. The output spectra represent the observed profiles in different regions and in  the observer's line of sight perpendicular to the slab.
} 
\label{scheme2}
\end{figure}

To understand these results, let us decipher the different
contributions in the simulation. 
Figure \ref{scheme1} depicts the situation for this
purpose. It shows the observed spectra in a simulation using
a point source and isotropic emission behind a uniform slab of
neutral gas. Observing the slab face-on, toward the source, we obtain
an absorption profile. Only photons far from the line center 
are transmitted directly, forming the ``continuum''. Photons in the line center
are resonantly absorbed and reemitted, diffusing in frequency and in
space, and leading to the lack of emission at and around the line
core. These photons will be collected if we look at the cube at a
position far from the source. A double-peak profile is then observed
consisting of the diffused photons and the absence of photons that
would have escape directly, without scattering, in this direction.

Figure \ref{scheme2} now shows a combination
of these single sources but with different intensities, illustrating
the extended source simulation of Fig.\ \ref{spatial_variation}. As
for the single source, the spectrum of the central region shows a
typical damped absorption. 
At the positions of the fainter peripheral sources
two contributions lead to a narrower absorption
profile: a) The transmitted flux is fainter
than in the central region, and b) photons that have diffused from the brighter sources to escape
further (double-peak emission), contributing to "fill the wings". 
In the central region the direct transmission is stronger and the diffuse part is weaker.  
In this way spatial variations of the UV continuum
combined with the resonant
transport effects of \lya\ radiation, can explain qualitatively the
observed \lya\ profile variations in \izw.
The observed profile in slit 7 (Fig.\ \ref{apertures_b}), may even show a hint
of the predicted double peak profile in its red wing, although the S/N is quite
low in this region.

Now we discuss the effect of adding \lya\ line emission on top of the UV continuum
emission. Naively one could expect a very different behavior given the very large
\lya\ equivalent width of the source in the peripheral parts of the NW region
(cf.\  Fig.\ \ref{profile_1d}).
However, the final spectrum remains unchanged despite the high \lya\ equivalent width used. It appears that the photons emitted at the core of the line are either destroyed by dust (\taua = 0.5 here) or backscattered, and only photons with higher frequency shift diffuse and contribute to the double peak emission. Therefore, increasing \ewlya\ has no incidence on the output spectrum since with FWHM(\lya)=100 \kms, all photons are emitted close to the center. This is easily confirmed by looking at the reflected spectrum (cf. Fig.\ \ref{scheme1}) which increases with higher \ewlya. We need to use unreasonably high FWHM(\lya) ($\ga$ 1000 \kms) to affect our result and see the double peak contribution increasing (in the profile wings). 
 This implies in particular also that our model predictions are insensitive 
to the observed spatial variations of \ewlya\ (cf.\ Fig.\ \ref{profile_1d}).

In short, we conclude that the observed variations of the \lya\ profile
across the NW region can be understood by a combination of the line of sight effects 
discussed earlier and by radiation transfer effects related to an extended source.

\subsection{Discussion}
\label{discussions}

For the first attempt to reproduce the damped absorption profile of \izw, we used a simple expanding shell model 
(Sect.\ \ref{shell_sec}).
If line-of-sight arguments could not be invoked, we would need
a relatively high column density (\nh\ = 6.5 $\times$ 10$^{21}$ cm$^{-2}$)  and a minimum amount of dust ($E(B-V)$ $\sim$ 0.05), which in this case is the only way to lose \lya\ photons. However, when we spatially select photons in the observer's sight-line, we showed (Fig.\ \ref{lya_fit2}) that one may observe \lya\ in absorption even without any dust ($E(B-V)$ = 0). These conclusions also hold for the SE region of \izw\ for which the integrated spectrum show a slightly larger \lya\ absorption (a Voigt fit yields \nh\ $\sim$ 4 $\times$ 10$^{21}$ cm$^{-2}$).       

\citet{martin96} found evidences of supergiant shell in \izw\ expanding at a speed of 35--60 \kms. The geometry proposed is a bipolar shell seen almost perpendicularly to its main axis (cf. their Fig.\ 4). This configuration is comparable to the shell geometry adopted here (Sect.\ \ref{shell_sec}) given the negligible effects of such small expansion velocities on our model spectrum. However, the output spectrum of the shell model would be significantly affected if  the \hi\ coverage is inhomogeneous and low column densities are observed in some sight-lines, which is still unclear here. For same reasons (low velocity and large \hi\ coverage), applying our extended geometry scenario (Sect.\ \ref{sight_line_sec}) to this configuration would yield same results, as our sight-line selection is still compatible with this ISM morphology.

 From our shell model we derived an \hi\ column density of $6.5 \times 10^{21} cm^{-2}$ which is higher than independant measurements of \nhi $\sim 2.6 \times 10^{21}$ \cm2\ from radio data by \citet{vanzee98}. Since, radio observations measure the total \hi\ content, our model would imply, for symetrical reasons, a value twice higher than given. However, with a typical beam size of 5\arcsec, the radio observations are not able to resolve the potential subparsec-size \hi\ clumps in the NW region, and the smoothing effects could lead easily to an under-estimation by a factor of $2-4$. Therefore, in absence of higher resolution observations, we can not rule out the supershell geometry.

The second model adopted led to a good fit with \nhi $= 3 \times 10^{21}$ and without dust comparable to observational constraints, and also explained the spatial variations of the absorption profile. It is therefore more likley that if the emission region is embedded in an \hi\ region, the geometry would be not symetric, with a higher column density in the front and/or ionised holes in the backside. Finally, it is worth noting that the geometry proposed by \citet{martin96} is not an embedded-like source and the expanding shell is bipolar and asymmetric with an axis inclined by $i \sim 10^{\circ}$ to our line of sight (see their Fig.\ 4), while \cite{vanzee98} find a higher inclination of $i \sim 55^{\circ}$.

\section{Comparison of \izw\ with other nearby and high-$z$ starbursts}
\label{comparison_sec}

We have just shown how with a low extinction or even no dust at all
it is possible to explain by radiation transfer and geometrical effects
the transformation of a strong intrinsic \lya\ emission
into the broad \lya\ absorption profile observed in \izw.
We need now to understand whether this galaxy is unique or representative of a certain class of objects and what our results
imply for other studies, including in particular \lya\ observations
of high-$z$ objects.

\subsection{Comparison with local starbursts}
Four of the eight \hii\ galaxies observed with GHRS/HST by
\citet{kunth98} show broad \lya\ absorption profiles: II Zw 70, Mrk
36, SBS 0335-052, and \izw\ studied here. As already noted by these
authors, these objects clearly differ from those with \lya\ in
emission by very low velocity shifts between the interstellar absorption
lines and the systemic velocity\footnote{Two of the objects with \lya\ in 
absorption, \izw\ and SBS 0335-052, have also been observed with FUSE.
with a large aperture. The measurements of \citet{grimes08} confirm
the earlier finding of low velocity shifts, now also on a much
larger aperture.}. An essentially static ISM appears therefore as
one of the main factors leading to \lya\ absorption, as already
concluded by these authors and as supported by our radiation transfer
modeling.

Furthermore, among the \lya\ absorbers, SBS 0335-052 and II Zw 70 show
clearly higher extinction, with $E(B-V) = 0.18$ and 0.15 respectively (less than 0.02 for Mrk 36, \citet{izotov98}). Hence the ISM properties of these objects 
should fulfill the same conditions, which have allowed us to explain
the \lya\ absorption of \izw, and dust destruction of \lya\ photons 
should be equally or more important.
Although very likely, we cannot fully prove this until \hi\ column density measurements
from the radio are available for all of them.
For SBS 0335-052 \nh\ reaches high values, up to $\la 9.4\times 10^{20}$ \cm2\
\citep{Pustilnik01}. Similarly, Mrk 36 shows a high column density peak up to $2.4 \times 10^{21}$ \cm2\ \citep{bravo04}.
Of course, depending on the efficiency of dust destruction, some spatial regions 
with diffuse \lya\ emission may be expected; however, this is not necessarily the case.
For example, for SBS 0335-052 we know that \lya\ absorption is observed 
over a large area, showing that absorption by dust must be important \citep{atek08}.

The other half of the HST sample of \citet{kunth98} shows \lya\ profiles in emission
and varying amounts of dust ($E(B-V)$ ranging from $\sim$ 0.02 to 0.18).
As already mentioned by these authors, the main difference with the other part of the sample
showing \lya\ absorption appears to be the clear signature of ISM outflows in the former.
A continuity of ISM velocities between ``static'' and ``outflowing'' is expected 
and observed \citep[see e.g.][]{martin05,grimes08}, mostly correlated with galaxy
luminosity, stellar mass, and SFR.
A more detailed analysis of the full sample of nearby starbursts observed in \lya\
will be presented elsewhere \citep{atek09}.

\subsection{Comparison with distant galaxies}
Compared to distant galaxies it is clear that \izw\ and SBS 0335-052, or at least
the regions of these objects showing intense star formation, show
very high \hi\ column density. For example, with $\nh \sim (0.9-3) \times 10^{21}$
\cm2, these two regions would correspond to the high \nhi\ tail of all DLA systems 
found in the SDSS DR3 survey 
\citep[cf.\ ][]{prochaska05}

Also, few high-$z$ starbursts with \lya\ absorption as broad as in \izw\
and SBS 0335-052 are known. While $\sim$ 25\% of the LBGs of \citet{shapley03}
show \lya\ absorption, their stacked spectrum shows a narrower absorption profile. 
Among the broadest \lya\ profiles of $z \ga 3$ LBGs are the two lensed
galaxies MS 1512--cB58 and FORJ0332-3557, whose
absorption profiles corresponds to $\sim (0.7-2.5)\times 10^{21}$
\cm2 \citep{pettini00,cabanac08}.

However, LBGs in general and these two objects in particular differ in many
properties compared to \izw. The objects with strong \lya\ absorption
show significant dust extinction ($E_\star(B-V) \sim 0.169 \pm 0.006$, where
$E_\star(B-V)$ is the color excess determined from stellar light), and high Star Formation Rate (SFR)
(dust-corrected SFR $\sim 52 \pm 5$ \myr). Furthermore outflows with significant
velocities ($\vexp \sim 100-300$ \kms) are generally observed in LBGs.
In comparison, \izw\ is a very low luminosity, low SFR object (with a UV luminosity
lower than that of LBGs by 2--3 orders of magnitude, SFR(UV) $\sim$ 0.3 \msolyr\ \citep{grimes08}
with a low extinction ($E(B-V) \la 0.05$), 
which shows a static ISM.

For LBGs \citet{schaerer08} have shown with radiation transfer models that 
the absorption profile of MS 1512--cB58 is due to the relatively large amount of 
dust and the high column density; with the observed ISM conditions this suffices
to transform intrinsic \lya\ emission expected from the ongoing starburst to
broad \lya\ absorption, despite the relatively large outflow velocity ($\vexp \sim
220$ \kms).
In \izw\ geometrical effects or a static high \nh\ ISM with small amounts of dust 
are sufficient to do a similar ``transformation''.

In short, we suggest schematically the following two explanations for \lya\ absorption
in nearby and distant starbursts:
{\em 1)} On average the cold ISM (relevant for \lya\ transfer) of LBGs shows
the geometry of a spherically expanding shell with relatively large velocities
and small variations ($\vexp \sim$ 100--300 \kms) \citep[cf.][]{shapley03,schaerer08,
verhamme08}. In such cases the main factors determining the escape fraction
of \lya\ photons are \nh\ and \taua, as shown by radiation transfer models
\citep{verhamme08,hayes09}, and significant amounts of dust are required
to obtain broad \lya\ absorption profiles.
{\em 2)} In nearby galaxies, small amounts of dust in a static/low velocity
ISM with a high \hi\ column density suffice to create \lya\ absorption. Furthermore, 
the occurrence of \lya\ absorption is most probably metallicity independent, at least
to first order. In addition, geometrical effects due to small apertures may also
increase the observed \lya\ absorption.

The distinction between groups 1) and 2) is most likely simply due to the outflow properties,
i.e.\ the wind velocity, which is known to increase with SFR, galaxy
mass, and specific star formation rate \citep[e.g.][]{martin05,rupke05,schwartz06,grimes08}.
Qualitatively this increase of the outflow velocity behavior is understood by
increasing mechanical feedback on the ISM related to stronger SF activity (SFR) in galaxies
with increasing mass or luminosity. At the low luminosity (SFR) end, feedback appears
to be insufficient to ``ignite'' outflows, hence the nearly static ISM in \izw\
and alike objects.
What ultimately settles the ISM geometry, \nh\ and dust to gas ratio, and 
hence assures in particular a high \hi\ column density in \izw\ and other local
objects remains to be explained.

Clearly, the observed trends and diversity need to be examined further both qualitatively
and quantitatively. This will be the scope of subsequent publications.

\section{Summary and conclusion}
\label{conclusions_sec}

Archival HST/STIS UV spectroscopy and imaging, and HST/WFPC2 optical imaging data of the nearby star forming galaxy \izw\ were obtained. We have applied the 3D Monte Carlo \lya\ radiative transfer code MCLya \citep{verhamme06} to explain quantitatively the intriguing \lya\ absorption in this galaxy and the apparent spatial variation of the \lya\ profile. Then, using the example of \izw, we have discussed under which physical conditions one observes \lya\ in emission or absorption both in nearby or high-z galaxies. Our main results can be summarized as follows:

\begin{itemize}
 \renewcommand{\labelitemi}{$\bullet$} 
 \item We first examined the predictions of a spherical shell model to reproduce the integrated spectrum of the NW region of \izw. This model described a static shell of \hi\ mixed with dust grains, surrounding a central point source emitting UV continuum plus a \lya\ emission line. Adopting dust extinction derived from observations ($E(B-V) = 0.05$) and $b=20$ \kms, we were able to fit the \lya\ profile with \hi\ column density of  \nh = 6.5 $\times$ 10$^{21}$ \cm2. Even with a strong intrinsic \lya\ emission \ewlya\ (up to 200 \AA) a
small amount of dust is sufficient to cause strong damped \lya\ emission, since the probability to be absorbed by dust is greatly increased by the high column density and by a nearly static ISM.
In this model, the loss of \lya\ photons is only possible by means of true dust absorption, since we spatially recover all photons.  
 
 \item Given the large spatial extension of \hi\ covering the NW region \citep{vanzee98}, we also explored the slab geometry of neutral gas in front of the UV source.  We have shown that considering only emission along the observer's line of sight, we can reproduce the strong \lya\ absorption without any dust ($E(B-V) = 0$). This is achieved by the diffusion of the photons out of the observer's sightline provided a sufficiently high column density (at least \nh = 3 $\times$ 10$^{21}$ \cm2) and a nearly static ISM configuration.
 
 \item We have observed spatial variations in the \lya\ profile shape in the different STIS slits. From observations we have constructed the 1D profile variations of the UV continuum and \ewlya\ across the NW region. Despite strong UV emission at the center and high \ewlya\ around, the \lya\ profile is still in absorption in all the NW region and proves narrower toward the peripheral region. 
To understand these variations we have simulated an extended source with a spatially varying UV emission strength by following the 1D spatial profile, in front of a slab of neutral gas with  \nh = 3 $\times$ 10$^{21}$ \cm2 and $E(B-V) = 0.05$. 
Then, by observing the slab at different distances from the center, we have been able to reproduce the observed spatial variations.
We have demonstrated qualitatively that this is due to radiative transfer effects, in particular, to the diffusion of \lya\ photons and to
the spatial variation of the UV continuum source.
Finally, adding a \lya\ recombination line to the source simulating the observed \ewlya\ profile, does not affect the final spectrum. 

\end{itemize}

Other nearby galaxies with intense star formation, such as II Zw 70, Mrk 36 and SBS 0335-052, also show strong \lya\ absorption \citep{kunth98}. They appear to show a nearly static cold ISM and more or less significant amounts of dust. 
At least two of these objects exhibit very high \hi\ column density (up to 2.4 $\times$ 10$^{21}$ \cm2).
We suggest that \lya\ absorption in these objects is due to the same reasons as for \izw: a very large number of scatterings in static, high column density gas leading to an efficient destruction of \lya\ photons by even small 
amounts of dust. Furthermore line-of-sight effects can also be responsible for or increase \lya\ absorption further.

The distinction between \lya\ emission and absorption in local starbursts seems to be mainly related to
presence or not of ISM outflows. Since high-$z$ objects (LBGs, LAEs) show generally outflows with high, 
but relatively similar velocities (with bulk velocity typically $\sim$ 100--200 \kms), the 
transition from \lya\ absorption to emission in these objects is, on the other hand mostly determined by 
the dust content and \hi\ column density \citep[cf.][]{schaerer08,verhamme08}.

These results and the global trends observed between \lya\ strength and profile diversity, and relevant parameters need 
now to be tested with larger samples of galaxies. This is the main objective of upcoming publications 
\citep[eg.\ ][]{atek09}.   


\begin{acknowledgements}
We would like to warmly thank Anne Verhamme, Matthew Hayes, Sally Heap and Fran\c{c}oise Combes for useful discussions. We are also grateful to Miguel Mas-Hesse for providing us with his UV spectra and for useful comments. 
Simulations were done on the {\tt regor} PC cluster at the Geneva
Observatory co-funded by grants to Georges Meynet, Daniel Pfenniger, and DS.
HA and DK are supported by the Centre National d'\'Etudes Spatiales (CNES). 
The work of DS is supported by the Swiss National Science Foundation.

\end{acknowledgements}

\bibliographystyle{aa}
\bibliography{references}

\end{document}